\documentclass[12pt]{article}
\usepackage{fullpage}
\usepackage{graphicx}
\usepackage{hyperref}
\usepackage{natbib}
\def\V{{\bf v}}
\def\B{{\bf B}}
\def\M{{\bf M}}
\def\x{{\bf x}}
\def\E{{\bf E}}
\def\K{\hat{\bf k}}
\def\L{\hat{\bf l}}
\def\J{\hat{\bf j}}

\def\ddtp{{\left.\ \ d\over c_s dt\right|_+}}
\def\ddtm{{\left.\ \ d\over c_s dt\right|_-}}
\def\ddtpn{{\left.d\over dt\right|_+}}
\def\ddtmn{{\left.d\over dt\right|_-}}
\def\gesim{\ \hbox to 0 pt{\raise .6ex\hbox{$>$}} \lower .5ex\hbox{$\sim$}\ }
\def\lesim{\ \hbox to 0 pt{\raise .6ex\hbox{$<$}} \lower .5ex\hbox{$\sim$}\ }

\begin{document}
\title{Oblique ion collection in the drift-approximation: how
  magnetized Mach-probes really work.}

\author{I H Hutchinson\\
Plasma Science and Fusion Center and \\
Department of Nuclear Science and Engineering,\\
 Massachusetts Institute of Technology,\\ Cambridge, MA 02139, USA}

\maketitle

\begin{abstract}
The anisotropic fluid equations governing a frictionless
obliquely-flowing plasma around an essentially arbitrarily shaped
three-dimensional ion-absorbing object in a strong magnetic field are solved
analytically in the quasi-neutral drift-approximation, neglecting parallel
temperature gradients. The effects of transverse displacements traversing the
magnetic presheath are also quantified. It is shown that the parallel
collection flux density dependence upon external Mach-number is
$n_\infty c_s \exp[-1 -(M_{\parallel\infty}- M_\perp\cot\theta)]$ where
$\theta$ is the angle (in the plane of field and drift velocity) of
the object-surface to the magnetic-field and $M_{\parallel\infty}$ is
the external parallel flow. The perpendicular drift,
$\M_\perp$, appearing here consists of the external $\E\wedge\B$ drift plus a
weighted sum of the ion and electron diamagnetic drifts
that depends upon the total angle of the surface to the magnetic field. It is
that somewhat counter-intuitive combination that an oblique
(transverse) Mach probe
experiment measures.

\end{abstract}

\section{Introduction}

Ion collection by solid objects immersed in a plasma is a problem of
perennial interest in plasma physics. It provides the basis for the
measurement of plasma parameters by electric (Langmuir)
probes\cite{HutchinsonBook2002} as well as the charging of
dust\cite{shukla02} and spacecraft\cite{hastings04}. The present
work addresses the situation where the ion Larmor radius (in the
background magnetic field ${\bf B}$) is much
smaller than the object, so that perpendicular plasma flow is strongly
constrained. 

This problem has important similarities to the solution of the flow of
plasma to a plane aligned obliquely to the field, the most obvious
example being a tokamak divertor plate. That problem can be
formulated\cite{riemann94} as one-dimensional, taking the coordinates
in the plane as being ignorable. However, it is well established that
no spatially-varying solution of the quasi-neutral plasma equations in
one dimension is possible without additional sources of particles
(e.g. through ionization) or momentum (e.g. from collisions). Recent
studies of this problem of the one-dimensional magnetized plasma and
oblique presheath (e.g. \cite{daube99,devaux06}) have mostly focussed
on collisions as the mechanism allowing the acceleration of the plasma
into the magnetic presheath. For localized probes, however, if one
conceptualizes the problem as being dominated by the one-dimensional
dynamics along the field, the \textit{cross-field flux divergence} is the
most natural effective source to permit parallel gradients.

Prior theoretical probe studies have focussed on situations where the
cross-field magnetized-ion flux can be described (somewhat
phenomenologically) as \emph{diffusive}. The full numerical solutions
for this formulation \cite{hutchinson87,Hutchinson1988} yield the
dependence of the collected ion flux density on the plasma density and
temperature, and the parallel (to ${\bf B}$) Mach-number. That
provides the theoretical calibration factor for a (parallel)
Mach-probe (a probe with electrodes facing parallel and anti-parallel
to the field), when the perpendicular drift velocity is ignorable. This
``calibration'' proves to be in good agreement with independent
measurements and calculations \cite{gunn01} and has been widely
adopted for experimental interpretation.

The approximate one-dimensional diffusive treatment has been
generalized\cite{VanGoubergen1999} to include an additional
perpendicular plasma drift velocity, accounting for the
boundary-condition modification\cite{Hutchinson1996} that the
transverse drift causes. Measuring the dependence of the ion
collection current-density on orientation of oblique probe faces then
allows one to deduce the perpendicular as well as the parallel external
drift velocity. The generalized solution can be
shown\cite{HutchinsonBook2002} to be a simple Galilean transformation of the
solution for zero transverse drift, which incidentally reminds us of
the elementary physical equivalence of ${\bf E}\wedge{\bf B}$ drift
past a fixed object and motion of the object through a stationary
plasma. The equivalence also indicates, though, that the generalized
diffusion solution is rigorously valid only for an oblique surface of
effectively infinite dimension in the transverse drift direction (so
that the Galilean equivalence in this direction is valid), but finite
in the direction perpendicular to both flow and magnetic field (so
that diffusion in this perpendicular direction dominates the cross-field
divergence). Practical Mach-probes generally do not have this
configuration. They are more often multi-faceted `Gundestrup'
types\cite{MacLatchy1992,gangadhara04,peleman06}, where many short adjacent
collectors are used with different orientations. So it is not obvious
that the generalized diffusive solution applies.

In fact, when there is substantial pre-existing cross-field drift of
the ions, it is perhaps physically more reasonable to regard that
drift as the dominant cross-field transport mechanism, and to
\textit{ignore diffusion}.  Gunn \cite{gunn98,gunn01} has explored
this problem, with a uniform impressed cross-field drift, in two
dimensions with his particle-in-cell code. This drift physics is appropriate
to many space and astrophysical problems too; for example to the
interaction of Jupiter's satellites with its magnetosphere.  It is
the purpose of the present work to derive a general analytic
(3-dimensional) solution to this purely advective problem, with fully
self-consistent drift velocity.

First, to introduce the solution by characteristics, we recall a recent
complete exact solution to this problem\cite{hutchinson08} for an
arbitrary shaped probe under the model ansatz that the perpendicular
drift velocity is uniform. This is a generalization of an earlier
self-similar solution\cite{hutchinsonreply88} mathematically
equivalent to a one-dimensional free expansion into a vacuum.  The
very simple general analytic result obtained for the ion flux is
gratifyingly close to the diffusive-plasma result, and hence to the
PIC calculations of Gunn (which include full ion distribution-function
parallel gradients). The solution demonstrates that provided the probe is
convex, the flux is not affected (for negligible Larmor radius) by
spatial derivatives of the surface angle in the drift-direction.  The
uniform drift ansatz is justified by inspection only when the probe
is two-dimensional; so that the coordinate perpendicular to the field
and drift is ignorable and the probe-perturbation of the plasma does
not introduce additional drifts except along the ignorable
coordinate. Again, practical probes are generally not
well-approximated as two-dimensional, so the question remains as to
whether that uniform-drift-velocity solution applies in practice.

The following remarkable result is rigorously demonstrated in section
\ref{drifts}. The ion flux to the probe surfaces derived for
uniform-drift-velocity \textit{does} apply even when the full
spatially-varying self-consistent drift velocity, including the
perturbation from an arbitrarily-shaped three-dimensional probe, is
accounted for. When the external drift arises purely from electric
field, one can obtain the full self-consistent spatial dependence of
the density and velocity throughout the perturbed plasma region, using
an elementary geometric algorithm. Some examples are given.

Furthermore, external \textit{diamagnetic drifts} can also be
included, again for arbitrary-shaped three-dimensional probes. They
make important but counterintuitive contributions to the observed ion
current density. In addition to the effects that arise in the plasma,
it is essential to account for transverse displacements that arise in
the magnetic presheath; they are calculated in section \ref{finite}.
Such local drifts in the magnetic presheath have previously been
identified\cite{chankin94,cohen95} as important contributors to
oblique boundary conditions. Based on these prior considerations, it
has sometimes been asserted that diamagnetic drifts do not transport
mass to a surface.  The present work provides a more general solution
of the magnetic presheath displacement effects, dispensing with
small-angle approximations, and clarifies rigorously the extent to
which diamagnetic drifts actually do contribute to the observed flux.

The final result is that a transverse Mach-probe measures effectively the
sum of the external ${\bf E}\wedge{\bf B}$ drift and a combination of
the ion and electron diamagnetic drifts. At small angles
between the field and the collector, the dominant diamagnetic term is
the \textit{electron} diamagnetic drift, which of course is generally
in the opposite direction to the ion diamagnetic drift. 

The presheath displacements can give rise to bias in Mach probe
measurements. Its relative magnitude is of order the ratio of Larmor radius
to electrode size. The effects of orthogonal displacements in the
plasma region are also calculated rigorously. They modify the
expression for the flux in ways that are usually of little importance
for practical measurements.

\section{Formulation}

We analyse the dynamics of the ion-fluid through the steady-state
continuity and momentum equations
\begin{eqnarray}\label{continuityeq}
  \nabla.(n\V) &=&0 \\
\label{momneq}
  mn(\V.\nabla)\V &=& -nZe\nabla\phi - \nabla p + n Ze (\V\wedge\B)\ ,
\end{eqnarray}
where $m$, $Z$, $n$, $p$, $T_i$, $\V$ are the ion mass, charge-number,
density, pressure, temperature, and velocity, and $\phi$ is the potential. 
We split the momentum equation into the components parallel ($_\parallel$) and
perpendicular ($_\perp$) to the (assumed uniform) magnetic field, and take the
cross-product with $\B$ of the perpendicular part to obtain the form
\begin{equation}\label{inclpol}
  \V_\perp = -\left[\left(\nabla_\perp\phi + {1\over
      nZe}\nabla_\perp p\right) + {m\over Ze} (\V.\nabla)\V_\perp\right]
      \wedge{\B\over B^2}\ .
\end{equation}
We can immediately identify the first two terms in this expression as
the ${\bf E}\wedge\B$ and diamagnetic drifts. The last term can be
considered to be the polarization drift, which we will regard as
ignorable. The approximation of omitting the polarization drift
requires the Larmor radius to be small c.f.\ the perpendicular
scale-length, generally the probe dimensions. It can be shown by
\textit{a posteriori} calculation that the polarization drift is smaller than
the imposed perpendicular drift by a factor that is second-order in
the Larmor radius. Ignoring the polarization drift term is the meaning
here of the expression ``drift-approximation''. By taking $\B$ to be
uniform we have of course eliminated the grad-B and curvature
drifts. We adopt the simplest possible fluid closure scheme, that the
ion temperature, $T_i$, is invariant, so that the pressure is simply
proportional to density. Together with dropping the polarization
drift, this makes $\V_\perp$ divergenceless:
\begin{equation}\label{drift_eq}
  \V_\perp = -\left(\nabla_\perp\phi + {1\over
      nZe}\nabla_\perp p\right)
      \wedge{\B\over B^2}
= -\nabla_\perp\left(\phi + {T_i\over Ze} \ln n\right)\wedge{\B\over B^2}
\ .  
\end{equation}
Under these approximations the continuity and
parallel-momentum equations become
\begin{eqnarray}
  (\V.\nabla)\ln n + \nabla_\parallel v_\parallel &=& 0 \\
\label{parmomen}  (\V.\nabla)v_\parallel + {Ze\over m}\nabla_\parallel\left(\phi +
  {T_i\over Ze}\ln n\right) &=& 0\ ,
\end{eqnarray}
while the perpendicular-momentum conservation is expressed by the
drift $\V_\perp$ expression. 

The potential is eliminated from these equations by accounting for the
self-consistent solution of the electric field arising from the ion
and electron densities. The electron density response along the
magnetic field is, as usual, taken to involve rapid equilibration; so
that the electron pressure gradient is balanced by electric field:
\begin{equation}
  \nabla_\parallel \phi = (T_e/e) \nabla_\parallel \ln n_e\ ,
\end{equation}
where the parallel gradient of the electron temperature,
$\nabla_\parallel T_e$, is taken to be zero and the electron density
is $n_e$. Assuming that the Debye length is much smaller than the
probe, we will treat the plasma as quasi-neutral, so that $n_e=Zn$.

Using the notation $c_s^2\equiv (ZT_e+T_i)/m$
and $\M\equiv\V/c_s$, the ion equations then take the normalized form:
\begin{eqnarray}
  \M.\nabla \ln n + \nabla_\parallel M_\parallel &=& 0 \\
  \M.\nabla M_\parallel + \nabla_\parallel \ln n &=&0\ ,
\end{eqnarray}
which can be rearranged by
adding and subtracting to show explicitly the ``characteristics'' \cite{courant62}
\begin{eqnarray}
\label{char_eqs1}
  (\M.\nabla + \nabla_\parallel)(\ln n + M_\parallel) &=& 0\\
\label{char_eqs2}
  (\M.\nabla - \nabla_\parallel)(\ln n - M_\parallel) &=& 0\ .
\end{eqnarray}
Thus the quantities $(\ln n \pm M_\parallel)$ are constant along their
corresponding characteristics $d\x=(\M\pm\B/B)ds$. And we can fully
solve the problem by analysis of the characteristics.  

\section{Uniform perpendicular-velocity ansatz}

First we review the solution under the condition that the
perpendicular velocity is simply a constant, $\M_\perp$, independent
of space. This ansatz is clearly justified if the coordinate perpendicular to
$\B$ and  $\M_\perp$ is ignorable. See ref \cite{hutchinson08}
for additional details and explanation of the following derivation. We
choose axes such that $\B$ is aligned along $x$ and $\M_\perp=M_h {\bf
  \hat{y}}$ along $y$. The requirements expressed by the
characteristics (\ref{char_eqs1}, \ref{char_eqs2}) are that both
\begin{eqnarray}
(\ln n +M_\parallel)&=&const\mbox{\quad along }dx=dy (M_\parallel+1)/M_h ,\label{positive}\\
\noalign{\noindent and}
(\ln n -M_\parallel)&=&const\mbox{\quad along }dx=dy (M_\parallel-1)/M_h .\label{negative} 
\end{eqnarray}
These will
be referred to respectively as the positive and negative
characteristics.

\begin{figure}[ht]
\medskip
\includegraphics[width=.6\hsize]{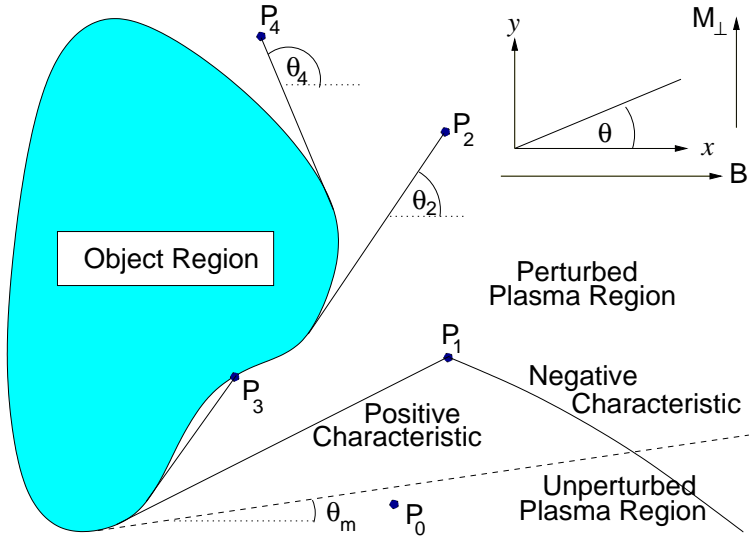}
\caption{\label{contsoln} (Color online). Construction of the
  solutions at different points. $P_0$ is in the unperturbed
  region. For $P_1$ the characteristics are shown. $P_3$ is in a
  concave region and so its positive characteristic is not tangent at
  $P_3$. A value $\theta > \pi/2$ such as for $P_4$ is not
  problematic.  }
\end{figure}

For definiteness, we now consider plasma that is on the higher-$x$
side (to the right) of the object. Figure \ref{contsoln} will be used
for illustration. For any point in the plasma, a positive and a
negative characteristic pass through it. If both characteristics
originate in the unperturbed plasma at $y\to-\infty$, and do not
enclose the object, then values at
the point satisfy both $\ln n + M_\parallel=\ln n_\infty+M_\infty$ and
$\ln n - M_\parallel=\ln n_\infty -M_{\parallel\infty}$ (where
$_\infty$ indicates values in the unperturbed region). These
simultaneous equations have only one solution: $n=n_\infty$,
$M_\parallel=M_{\parallel\infty}$, showing that the point is in the
unperturbed region, for example: $P_0$. The characteristics for such
points are straight lines with slopes $M_h/(M_{\parallel\infty}\pm
1)$.

The most important case is when just one of the characteristics
originates not at $y=-\infty$, but on the surface of the object
(e.g.\ P$_1$).  The positive characteristic is always to the left of
the negative characteristic, when tracing backward from a common
point.  So the characteristic that originates on the object is the
positive one. On that characteristic, $(\ln n + M_\parallel)$ is
constant, but not equal to the unperturbed value. Each point along the
positive characteristic satisfies also $\ln n - M_\parallel=\ln
n_\infty -M_{\parallel\infty}$ because there are negative
characteristics from infinity to each point. The only way to satisfy these two
requirements is that, along the positive characteristic,
$M_\parallel=const$ and $n=const$. If $M_\parallel=const$, then the
slope of the characteristic, $M_h/(M_\parallel+1)$ is constant. It is
a straight line.

The line's slope is determined by the absorbing boundary condition at
the plasma edge. That condition requires\cite{hutchinson08}
$M_\parallel$ to be as negative as possible consistent with the
overall solution, which requires the
greatest possible slope-angle $\theta\equiv\arctan[M_h/(M_\parallel+1)]$ (even
perhaps such that $\theta > \pi/2$).  The characteristic must
therefore always be \emph{tangential} to the object boundary where it
intersects it. Thus, all positive characteristics that originate on
the boundary do so as tangents, and for any point in the perturbed
plasma region \emph{the positive characteristic is that straight line
  passing through the point which has greatest $\theta$ and originates
  as a tangent on the object}. Once that line is determined
geometrically, its slope determines $M_\parallel$ and hence $n$ at all
points along it. If the steepest tangent angle is less than $ \theta_m
= \arctan[M_h/(M_{\parallel\infty}+1)]$, then the positive
characteristic does not intersect the object, but has slope
$\theta=\theta_m$; and the point is in unperturbed plasma. The entire
solution for the plasma in the perturbed neighborhood of an
arbitrary-shaped object is thus
\begin{equation}\label{curved}
  n=n_\infty \exp(M_\parallel-M_{\parallel\infty}),\quad M_\parallel=M_h \cot\theta -1.
\end{equation}

The solution (\ref{curved}) provides an extremely simple formula
for the ion flux to a surface not affected by concavity. Adding
the perpendicular and parallel components, the total flux density
along the outward normal within the $(x,y)$-plane, i.e.\ in
the direction $(-\sin\theta, \cos\theta)$, is 
$n c_s(M_h \cos\theta - M_\parallel \sin\theta) = n c_s\sin\theta$. Written
as flux per unit area perpendicular to the magnetic field, this is
\begin{equation}\label{flux}
  \Gamma_\parallel = n_\infty c_s \exp[-1 -(M_{\parallel\infty}- M_h\cot\theta)].
\end{equation}

First, this form indicates, importantly, that for points not in a
concave region of the object, the collected flux depends only on the
angle of the surface there, and not on the shape of the object at
smaller $y$. Second, the exponential dependence upon
$M_{\parallel\infty}$ is within 10\% of the dependence,
$\exp[-1-1.1(M_{\parallel\infty} - M_h\cot\theta)]$, that fits the
diffusive solution\cite{VanGoubergen1999,HutchinsonBook2002}. Third,
consideration of the characteristics shows unambiguously that leading
faces, for which $\theta<\theta_m$ receive simply the unperturbed flux
[$\Gamma_\parallel = n_\infty c_s (M_h \cot\theta -
  M_{\parallel\infty})$], while trailing faces, even those for which
$M_{\parallel\infty}- M_h\cot\theta>1$, are governed by the formula
(\ref{flux}). The boundary condition at the magnetic presheath edge is
just the same as what is sometimes called the ``magnetized Bohm
condition'' but arises here naturally from the analysis of the
quasi-neutral equations.

In a concave region of the object, e.g. at $P_3$, the surface angle (local tangent)
$\theta_s$, is smaller than the characteristic's angle $\theta$, and the
distinction must be retained. This leads to an enhanced ion flux, equal to
equation (\ref{flux}) times the extra factor
$(M_h\cot\theta_s-M_h\cot\theta +1)$.

\section{Accounting for self-consistent drifts}\label{drifts}

The presence of the probe perturbs the plasma potential and density,
for example as calculated in the model case of the previous
section. Most probes are of limited extent in the direction ($z$)
perpendicular to the plane containing $\B$ and $\M_{\perp\infty}$, and
indeed may have a non-zero $z$-component of the normal to the
collecting surface.  The plasma perturbations therefore give rise to
spatially-varying ion drifts that frequently break the
($z$-translational) symmetry assumption used to justify the
homogeneous-$\M_\perp$ ansatz. So we must now return to the full
equations (\ref{char_eqs1}, \ref{char_eqs2}) accounting for the
complete, self-consistent, spatially-varying, perpendicular velocity.

The interesting case is of points for which (only) one of the
characteristics starts not at infinity but on the probe itself. As
before, for definiteness, but without loss of generality, we will take
that to be the positive characteristic. At this point then $(\ln n +
M_\parallel)\not=(\ln n_\infty + M_{\parallel\infty})$; nevertheless,
because of the negative characteristic, it is still true that $(\ln n
- M_\parallel)=(\ln n_\infty - M_{\parallel\infty})=const$. This shows
that in a region whose points have any characteristic starting at
infinity, the nature of the solution is of the form
$M_\parallel=M_\parallel(n)$, and especially $\nabla M_\parallel$ is
parallel to $\nabla n$. Consequently if there is a self-consistent
combination of density and velocity fields \{$n$, $M_\parallel$,
$\M_\perp=\M_{\perp0}$\} that satisfies the advection equations
(\ref{char_eqs1}, \ref{char_eqs2}) (and the drift equation,
\ref{drift_eq}), any perpendicular vector field, $\M_{\perp1}$, that
satisfies $\M_{\perp1}.\nabla n=0$, also satisfies $\M_{\perp1}.\nabla
M_\parallel=0$. We can therefore subtract any such $\M_{\perp1}$ from
$\M_\perp$ without affecting the characteristic equations
(\ref{char_eqs1}, \ref{char_eqs2}). In other words, the combination
\{$n$, $M_\parallel$, $\M_\perp=(\M_{\perp0} - \M_{\perp1}$)\}
\textit{also} satisfies the characteristic equations (though not the
drift equation, \ref{drift_eq}). Moreover, the subtraction of
$\M_{\perp1}$ leaves the boundary condition at the plasma edge
invariant provided probe curvature is small compared with
$1/\rho_s$. The invariance follows from the boundary condition
being that the positive characteristic be tangent to the surface. If the
surface is expressed by the equation $s(x,y,z)=0$, then tangency is
$d{\bf x} .\nabla s =0$, which along the characteristic is
$(M_\parallel+1)\nabla_\parallel s + \M_\perp.  \nabla_\perp s=0$. In
so far as $\nabla s$ and $\nabla n$ are parallel (i.e.~the density is
invariant in the tangential directions) subtracting $\M_{\perp 1}$
leaves this condition unchanged. But $n$ is indeed invariant along the
surface when given by eq (\ref{curved}) provided that the surface
angle ($\theta_s$) gradient (i.e.~curvature) can be ignored.

The significance of these observations is profound. It means that
although the actual perpendicular velocity $\M_\perp$ may be very
complicated, and include drifts arising from the self-consistent
potential- and density-gradients caused by the presence of the probe,
we do not have to solve the associated complicated
characteristics. Instead, we can subtract from $\M_\perp$ any drift
that satisfies $\M_{\perp1}.\nabla n =0$, and solve the resulting
simpler characteristics. The resulting solution for $n$ and
$M_\parallel$ is correct then also for the full drift $\M_\perp$
expression. 

\subsection{Pure \texorpdfstring{$\E\wedge\B$}{ExB} external drift}

Consider first the case when the unperturbed plasma has uniform
density ($n_\infty= const$) as well as temperature, but has a uniform
impressed perpendicular electric field in the z-direction
perpendicular to the field , ($\B/B = {\bf \hat x}$), giving rise to
an $\E\wedge\B$ drift. In other words, we have a non-uniform potential,
$\phi_\infty(z)$, such that $\nabla \phi_\infty = -{\bf \hat z}E$; so
that the homogenous drift is $\V_h= {\bf \hat y} E/B = \M_h c_s$.

In the presence of the probe, the electron parallel momentum
(force-balance) equation can be integrated along the field, from
infinity to any position to give the perturbed potential
\begin{equation}\label{phiint}
  \phi - \phi_\infty(z) = (T_e/e)\ln(n/n_\infty)\ .
\end{equation}
We substitute this into the drift expression to get
\begin{eqnarray}\label{phisub1}
  \V_\perp &=& - \nabla \left(\phi_\infty + {T_e\over e}\ln(n/n_\infty)
  + {T_i\over Ze} \ln n\right)
      \wedge{\B\over B^2} \\
\label{phisub2}
   &=& -\left(\nabla \phi_\infty + {m c_s^2\over Ze} \nabla \ln
      n\right)      \wedge{\B\over B^2}\ .
\end{eqnarray}
Or, in normalized form:
\begin{equation}
  \M_\perp = \M_h - \rho_s \nabla \ln n \wedge\B/B\ ,
\end{equation}
where $\rho_s$ is the ion Larmor radius at the sound speed.  The
perpendicular drift velocity thus consists of a uniform term $\M_h$,
equal to the external drift, plus a term $\M_{\perp1}$ ($=-\rho_s
\nabla \ln n \wedge\B/B$) arising from local gradients of density (and
associated potential), which satisfies $\M_{\perp1}.\nabla n =0$. Our
approach is therefore to solve along characteristics defined not by the
complicated full drift velocity $\M_\perp$, but using the uniform
external drift velocity $\M_h$ which arises from subtracting off
$\M_{\perp1}$, as we have shown we are permitted to do because
$\M_{\perp1}.\nabla n =0$.  But such an approach, of using a uniform
impressed perpendicular drift, is precisely the ansatz solved in the
previous section, albeit without this detailed justification.
Therefore, the solution obtained there applies without modification.
What \textit{is} modified is that the condition of translational
invariance in the orthogonal (z-) direction is removed. In other
words, the restriction that the probe be two-dimensional, which was
invoked previously to justify the neglect of the
density-gradient-induced drifts is proven here to be unnecessary.  The
results of that section, the dependence of $n$ on $M_\parallel$, the
spatial variation of $M_\parallel$ embodied in the equation
$\tan\theta=M_h/(M_\parallel+1)$, and the surface flux expression,
apply to any shape of three-dimensional probe, provided only that the
condition of convexity (that the surface is not reentrant) is
satisfied. This convexity condition must be applied along the
original characteristic (before subtracting $\M_{\perp\infty}$) so it
depends on the full 3-D shape of the probe (and the $z$-drifts).

The flux-density of ions from the plasma to the probe surface in this case
is given by the vector sum of $\M_h$ and $M_\parallel$ because the
additional term $\M_{\perp1}$ is always locally tangential to the
surface, even if the surface-normal has a non-zero
$z$-component. Therefore eq (\ref{flux}) is valid.

\begin{figure}[htp]
  \includegraphics[width=.7 \hsize]{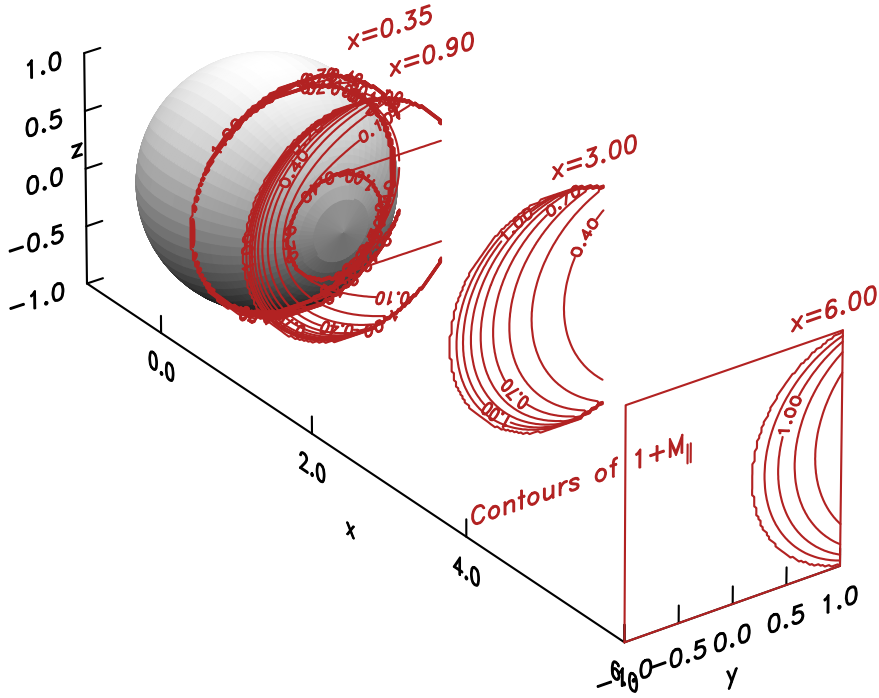}
  \caption{\label{sphere} (Color online). Contours of
    $M_\parallel+1=1+M_{\parallel\infty} + \ln(n/n_\infty)$ at intervals of
    0.1, in planes of
    constant-$x$ near a sphere of unit radius in a plasma with
    external drift velocity $M_{\parallel\infty}=0.2$, $M_h=0.24$.}
\end{figure}

\begin{figure}[htp]
  \includegraphics[width=.7 \hsize]{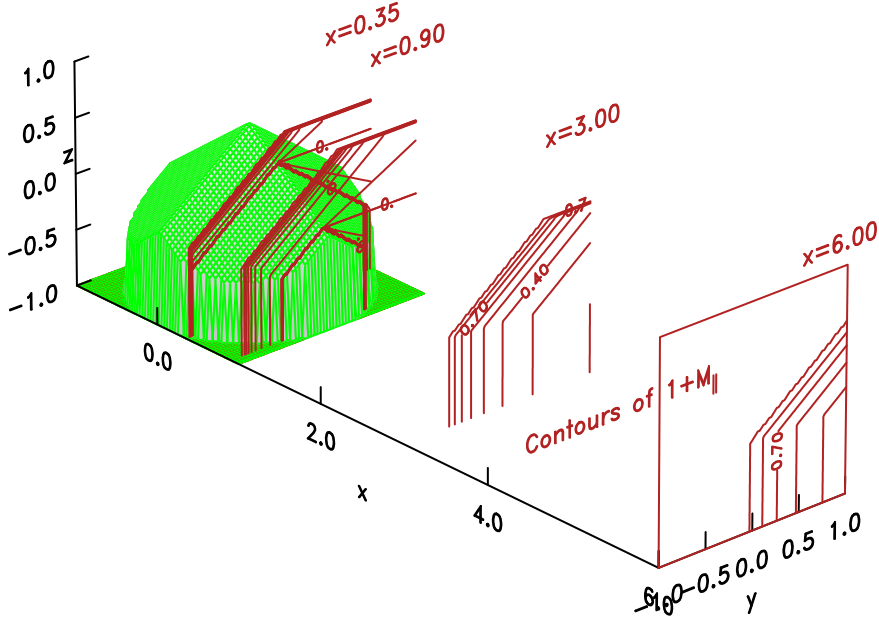}
  \caption{\label{pyramid} (Color online). Contours of
    $M_\parallel+1=1+M_{\parallel\infty} + \ln(n/n_\infty)$ at intervals of
    0.1, in planes of
    constant-$x$ near a pyramidal Mach probe in a plasma with
    external drift velocity $M_{\parallel\infty}=-0.1$, $M_h=0.15$.}
\end{figure}

As an illustration of this complete solution of the problem, Fig
\ref{sphere} shows a representation of the 3-Dimensional variation of
$M_\parallel$ and equivalently $\ln n$ by contours of the quantity
$M_\parallel+1$ drawn in perpendicular planes at various distances
from a spherical object. Regions empty of contours to lower-$y$ (left)
of the contours shown have uniform unperturbed plasma. To the right of
the contours is the wake region where the equations are not valid.

A second illustration is in Fig \ref{pyramid}, which shows the
contours for a pyramid shaped probe similar to what is used in Alcator C-Mod
experiments\cite{smick06}. A different drift velocity is illustrated.

\subsection{Inclusion of external \texorpdfstring{$\nabla n$}{grad-n} diamagnetic drift}

The presence of a diamagnetic drift arising from external density
gradient leads to several complicating factors. We consider an
unperturbed density, $n_\infty$, that in this case is not uniform but
is a function also of perpendicular position. The usual case has $\nabla_\perp \phi_\infty$ and $\nabla_\perp
n_\infty$ approximately parallel to each other, because both are
perpendicular to the flux surface in a confined plasma. This is actually
essential for the consistency of the unperturbed state, if it has no
parallel gradients. Then the
total drift velocity does not satisfy $\nabla. (n_\infty \V_\infty)=0$
unless $\B.(\nabla \phi_\infty\wedge\nabla n_\infty)=0$ so that the
drifts arising from the $\phi$ and $n$ gradients are parallel to each other.
We stick to this case here, and consistently take all external
gradients to be in the $z$-direction, but note that
effects in tokamak scrape-off-layers where parallel gradients are present
and give rise to cross-flux-surface drifts are thereby ignored. In
other words, we are dealing with a case of negligible parallel
gradients in the external plasma.

We suppose that the external logarithmic gradient of the density is
constant:
\begin{equation}
  \nabla \ln n_\infty(z) = \hat{\bf z} /L_n ,
\end{equation}
where $L_n$ is the (constant) density scale length.
Equations
(\ref{phiint}) and (\ref{phisub1}) still apply. We still can rely upon
integration along the negative characteristic to write down a
relationship between $n$ and $M_\parallel$, but that relationship is
now
\begin{equation}
  \ln n - M_\parallel = \ln (n_\infty(z_\infty)) - M_{\parallel\infty} ,
\end{equation}
where $n_\infty(z_\infty)$ is the unperturbed value of density at
large distance from the probe, but at a value, $z_\infty$, of $z$
corresponding to tracking backward along the negative characteristic
from the point of interest. Because $n_\infty$ is a function of $z$
when there are diamagnetic drifts, the value of $\ln
n_\infty(z_\infty)$ depends upon the total $z$-displacement, $\delta z
= z - z_\infty$ between the characteristic's start and the point of
interest. Thus it is no longer the case that $n=n(M_\parallel)$.

Write the relationship between density and $M_\parallel$, deduced from
the negative characteristic integration as
\begin{equation}\label{diaint}
  \ln (n/ n_\infty(z)) = -M_{\parallel\infty} + M_\parallel - \delta
  z/L_n .
\end{equation}
In this expression, $\delta z$ is not a constant. We will demonstrate
in the following, however, that a solution exists in which $\delta z$
is a function only of $M_\parallel$. So taking $\delta z = \delta
z(M_\parallel)$ we observe that $\ln(n/n_\infty)$ is also a function
only of $M_\parallel$. [Here and following we use the notation
  $n_\infty$ without an argument to denote the unperturbed density
  at the position of the point, $n_\infty(z)$ not
  $n_\infty(z_\infty)$.] 
The drift velocity from eq (\ref{phisub1}) can be written
\begin{equation}\label{gradndrift}
     \V_\perp= -\left[\nabla \left(\phi_\infty + {T_i\over Z e}\ln
       n_\infty\right) + {m c_s^2\over Ze} \nabla \ln (n/n_\infty)\right]      
\wedge{\B\over B^2}\ ,
\end{equation}
in which the first two terms give rise to perpendicular Mach number
$M_h=M_E+ M_{ni}$, the sum of external $E\wedge B$ and ion diamagnetic
drifts. The final term, which we identify as $\M_{\perp1}$ is
perpendicular to $\nabla \ln n/n_\infty$ and, because $\ln n/n_\infty$
is a function of $M_\parallel$, perpendicular also to $\nabla
M_\parallel$.

Now we write the positive
characteristic equation so as to use these expressions:
\begin{eqnarray}\label{diacharac}
  0 &=& \ddtp \left(\ln n + M_\parallel\right) =
 \ddtp \left(\ln n/n_\infty + M_\parallel\right) + M_z/L_n
\nonumber\\
  &=& M_z/ L_n + \ddtp \left(2 M_\parallel - \delta z /L_n\right)
  = M_z/ L_n + \ddtp \left(2 \ln n/n_\infty + \delta z /L_n\right) ,
\end{eqnarray}
where we use the notation
\begin{equation}
  {\left.\ \ d\over c_s dt\right|_\pm}\equiv \M.\nabla \pm
  \nabla_\parallel = 
    (M_\parallel \pm 1)
    {\partial\over \partial x} +
    M_y {\partial\over \partial y} +
    M_z {\partial\over \partial z} 
\end{equation}
for the derivative along the positive or negative characteristics.
Because the arguments of the derivatives in the last two forms of eq
(\ref{diacharac}) are
explicitly functions only of $M_\parallel$, we can subtract
$\M_{\perp1}.\nabla$ from the characteristic derivative without
effect. In other words, in those expressions, we can interpret the
derivative in the alternative version (valid only when operating on
functions only of $M_\parallel$)
\begin{equation}\label{reint}
  {\left.\ \ d\over c_s dt\right|_\pm}= (M_\parallel \pm 1)
    {\partial\over \partial x} +
    M_h {\partial\over \partial y}
\end{equation}
Now we eliminate $y$-derivatives of $\ln n$ from the last form of the positive
characteristic equation using the expression for the
$z$ drift velocity:
\begin{equation}\label{grady}
  M_z = \rho_s {\partial \ln n \over \partial y} = \rho_s {\partial
    \ln n/n_\infty \over \partial y}
\end{equation}
to find
\begin{equation}\label{gradx}
  (M_\parallel + 1) {\partial \over \partial x} \ln n/n_\infty = 
- M_z \left({1\over 2 L_n} + {M_h\over \rho_s}\right) - {1\over 2 } \ddtp
  {\delta z\over L_n} .
\end{equation}
In the same way, we express the negative characteristic equation in
terms of $\ln n/n_\infty$, reinterpret the derivative as the form (\ref{reint}) and 
then we eliminate $x$- and $y$-derivatives of $\ln n/n_\infty$ using eqs
(\ref{gradx}) and (\ref{grady}).
\begin{eqnarray}\label{dianeg}
  \ddtm M_\parallel &=& \ddtm \ln n/n_\infty + M_z/L_n \nonumber\\
  &=& {M_\parallel -1\over M_\parallel +1} \left[ 
- M_z \left({1\over 2 L_n} + {M_h\over \rho_s}\right) - \ddtp
  {\delta z\over 2 L_n}\right]+ {M_h M_z\over \rho_s} + {M_z\over L_n}
.
\end{eqnarray}
For compactness we define
$d/dM_\parallel(\delta z/2L_n) \equiv r$; so that 
\begin{equation}
  \ddtp{\delta z\over 2 L_n} = \ddtp M_\parallel . {d\over dM_\parallel}
{\delta z\over 2 L_n}
= r \ddtp M_\parallel .
\end{equation}
Eliminate $\ddtp M_\parallel$ between this expression and the
$M_\parallel$ form of eq (\ref{diacharac}) to obtain
\begin{equation}
  \ddtp{\delta z\over 2 L_n} = - {M_z\over 2 L_n} {r\over 1- r} .
\end{equation}
Substituting into eq (\ref{dianeg}) and eliminating $M_z$ and the right
hand side's $\ddtm M_\parallel$ using the identity
\begin{equation}
  r = \ddtm \left.{\delta z\over 2L_n}\right/ \ddtm M_\parallel 
    = {M_z\over 2L_n} \left/ \ddtm M_\parallel\right. ,
\end{equation}
we arrive at the following quadratic equation for $r$
\begin{equation}
  1 = {r\over M_\parallel + 1} \left[ M_\parallel + 3 + {r\over 1-r}
    (M_\parallel -1) + 4 L_n M_h/\rho_s\right] .
\end{equation}

We solve this equation, using the simplifying notation 
\begin{equation}
  a \equiv 1/(4 +4L_nM_h/\rho_s)\ ,\qquad u\equiv a M_\parallel\ ,
\end{equation}
to find
\begin{equation}\label{requ}
  r = (1/2)[1+2u \pm \sqrt{1-4a+4u^2}]\ .
\end{equation}
Then we can integrate to obtain the $z$-displacement
\def\quart{{\scriptstyle{1\over 4}}}
\begin{equation}
  {\delta z \over L_n} = 2\int r dM_\parallel = 
  {1\over a}\left[u + u^2 - u \sqrt{u^2-a+\quart} +
   (a-\quart) \ln\left(u +\sqrt{u^2-a+\quart}\right)\right]_{u_\infty}^u,
\end{equation}
 This solution is real only if $\quart-a > 0$, which is equivalent to
 $\rho_s/(L_nM_h) >-1$.  Since the ion diamagnetic drift is
 $M_{ni}=-\rho_s/L_n$ and $M_h$ is positive by presumption, this
 requires $M_{ni}>-M_E$ and $M_E>0$.

\begin{figure}[ht]
\medskip
\hskip 1em(a)\hskip-2.5em
\includegraphics[width=.5\hsize]{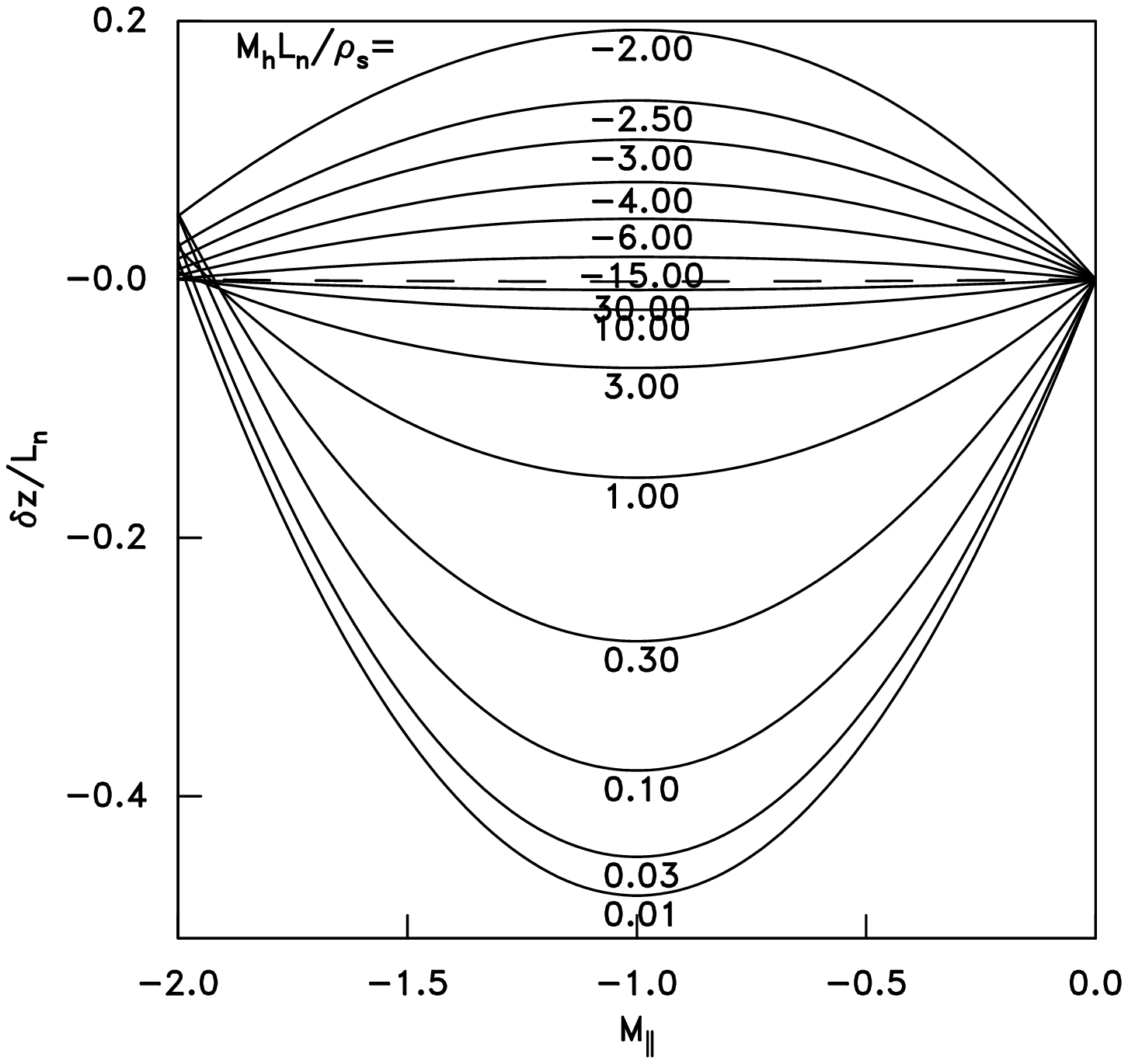}
\hskip 1em(b)\hskip-2.5em
\includegraphics[width=.5\hsize]{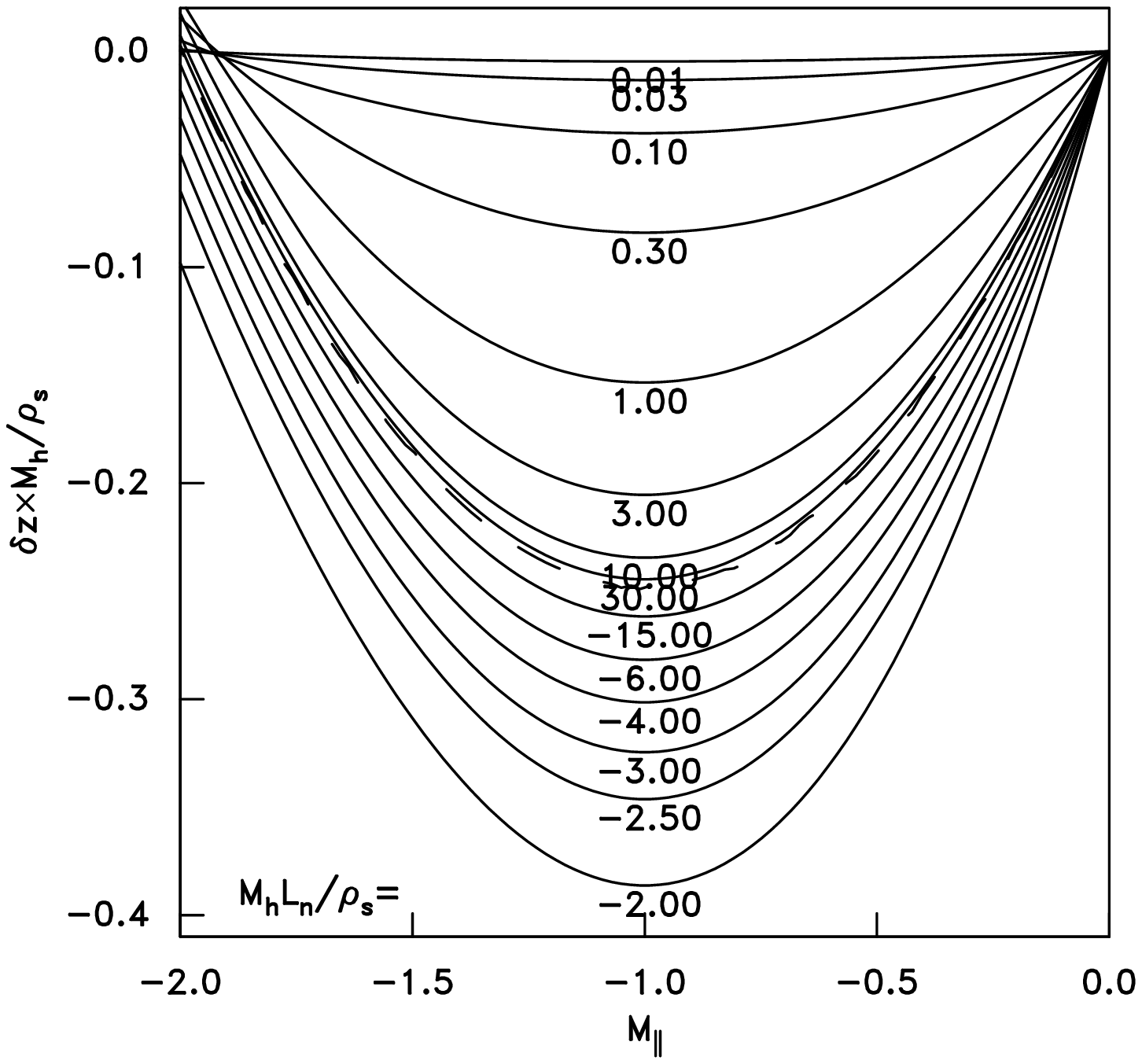}
\caption{\label{dzsoln}Solutions for the $z$-displacement normalized
  to the density scale-length (a) or the Larmor radius (b), for the
  useful range $-M_E<M_{ni}<M_E$. It is assumed that
  $M_{\parallel\infty}$ is zero, but the curve shapes are the same
  regardless of the starting value of $M_\parallel$.
}
\end{figure}

In Fig \ref{dzsoln} are plotted the solutions for $\delta z$ as a
function of $M_\parallel$ for the entire useful range of $a$, i.e. of
$M_hL_n/\rho_s = -M_h/M_{ni}$. We take that to be constrained by
$-M_E<M_{ni}<M_E$ (and $M_E>0$) which is equivalent to
$\rho_s/(L_nM_h) >-1/2$. Whether normalized by $L_n$ or by
$\rho_s/M_h$, these results show that the displacement, $\delta_z$, while
non-zero, is modest. Moreover the curves are quite close to being
mirror-symmetric about the line $M_\parallel=-1$.

The value of $\delta z$ together with the negative characteristic
integration, eq (\ref{diaint}) provide the complete solution for $\ln
(n/n_\infty)$ which is a function only of $M_\parallel$. The
characteristic equations can be considered to be 
\begin{equation}\label{ncharacteristics}
  {\left.\ \ d\over c_s dt\right|_\pm} (\ln n/n_\infty \pm M_\parallel)
  = - {M_z\over L_n},
\end{equation}
in which the characteristic derivatives can be taken as eq
(\ref{reint}), that is, lying in a $z=const$ plane.  The
characteristics are now curved, and cannot be constructed directly
from the geometry, as was possible in the absence of diamagnetic
drift. But this does not matter for the purposes of obtaining the flux
to the probe.
The boundary condition at the plasma edge is, as before, that provided
the boundary is convex, the positive characteristic must be tangential to
the surface; that is, $M_\parallel+1 = M_h \cot\theta$. 

In practice, the perpendicular velocity is generally deduced from Mach
probe measurements effectively by comparing values of the ion current
density for faces having equal and opposite values of $\cos\theta$;
that is, having the same angle to the magnetic field, but pointing
upstream or downstream with respect to the perpendicular
flow. Although $\delta z$ changes the value of $\ln n$ and thus
affects the ion flux density, if $\delta z$ is of exactly even parity
in $M_\parallel + 1$ and hence in $\cot\theta$, it will contribute
nothing to the ratio of the ion fluxes on which the measurement is
based. For $L_n\to \infty$ (zero diamagnetic drift) or $M_h L_n=0$
(zero net drift) $\delta z$ is indeed exactly of even parity. The
maximum value of the odd-parity part of $\delta z/L_n$ is
approximately 0.02 for $|M_\parallel+1|<0.6$ for
$|M_{ni}/M_E|\le 1$ (and rapidly decreasing as $|M_{ni}/M_E|$
decreases). Thus the contribution of the $\delta z$ term to Mach probe
velocity measurement is typically less than 2\%, which is for
practical purposes negligible. The magnitude of $\delta z$ is
sufficiently great, however, that it would be inadvisable to attempt
to deduce the perpendicular flow without taking advantage of its
approximate parity. In other words, a Mach-probe measurement really
must be based on electrodes with equal and opposite values of
$\cot\theta$, and not, for example, on comparing positive $\cot\theta$
with $\cot\theta=0$.

\begin{figure}
  \hskip 10em(a)\hskip-11.5em
\includegraphics[width=.5\hsize]{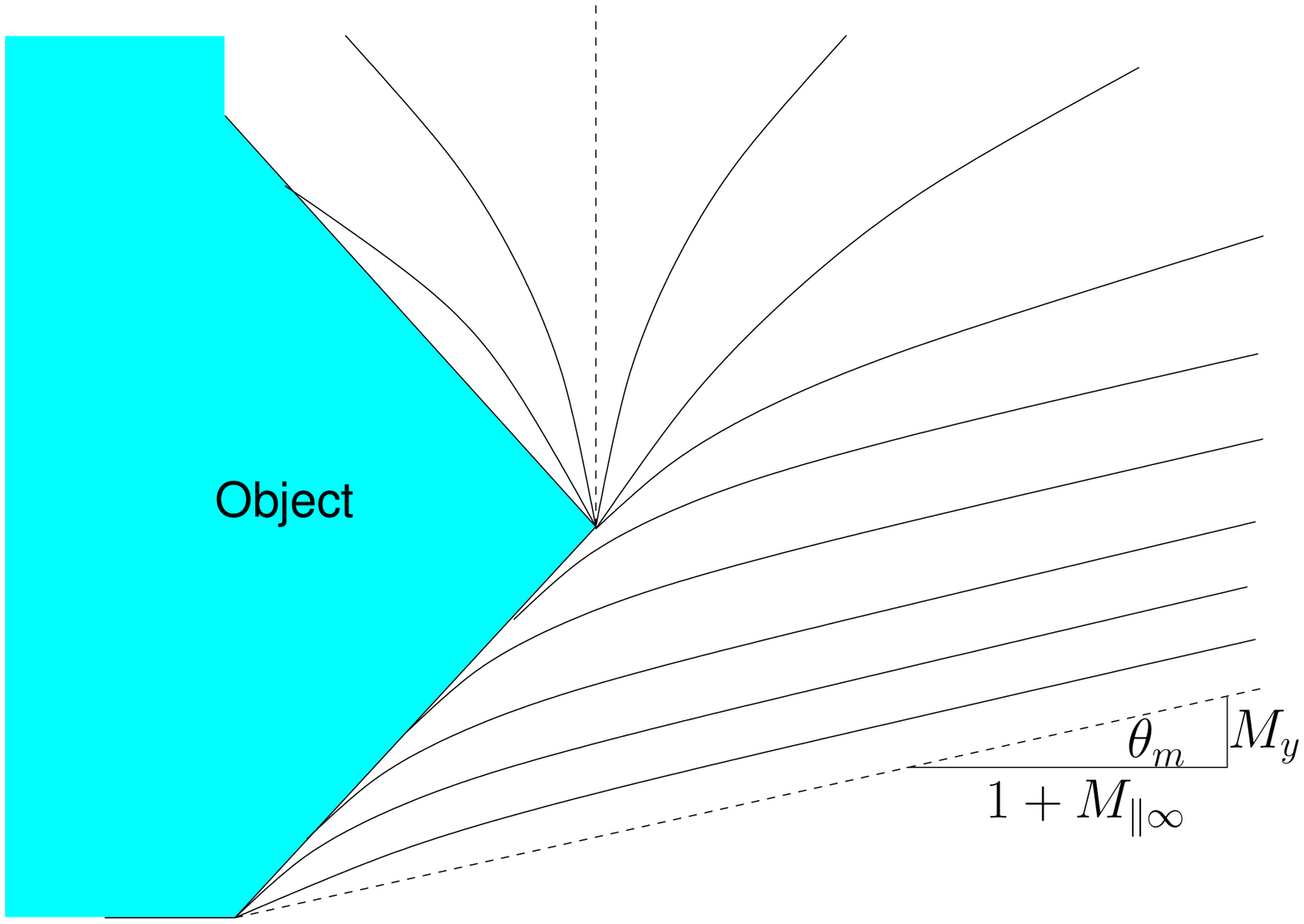}
\hskip 10em(b)\hskip-11.5em
\includegraphics[width=.5\hsize]{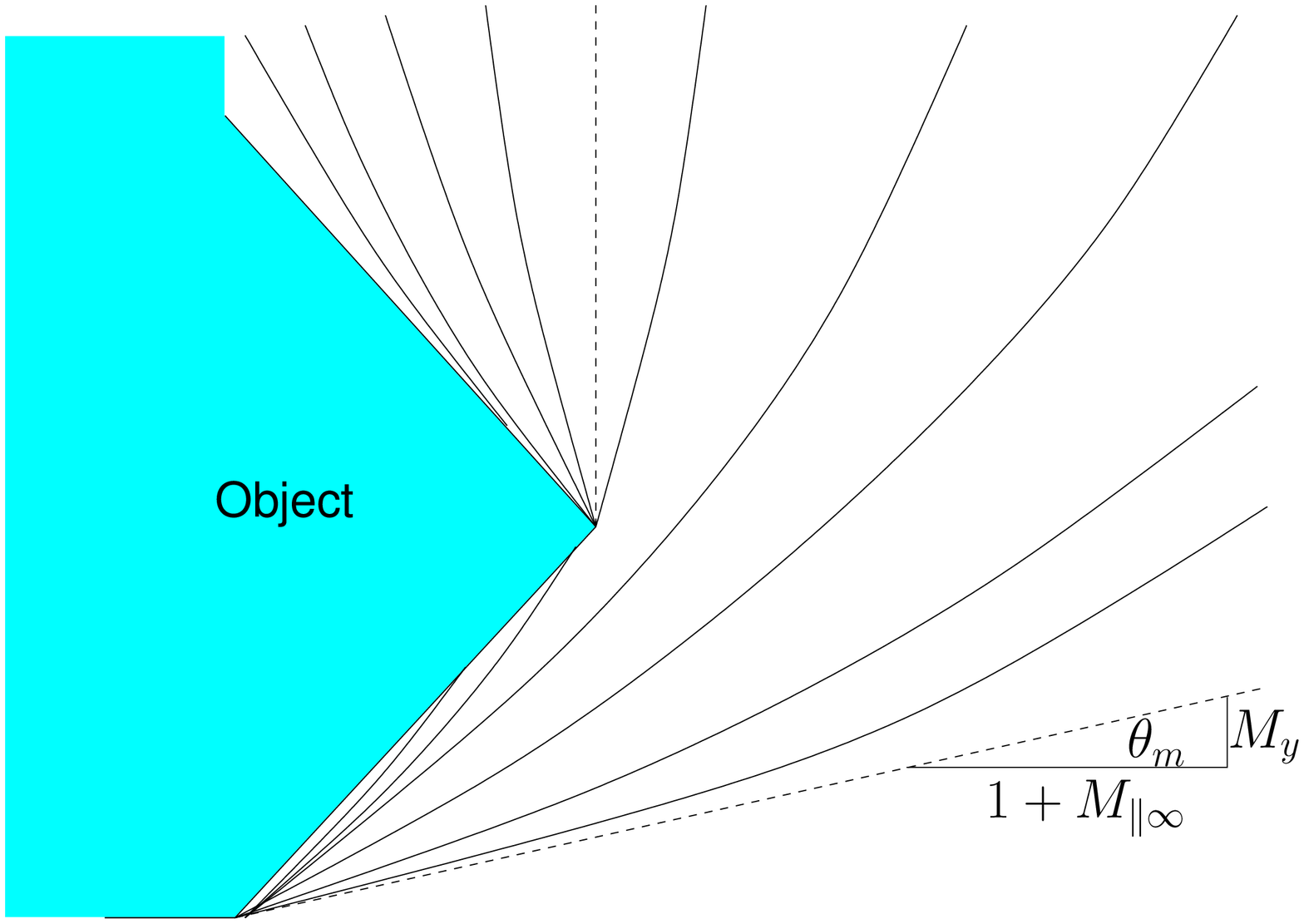}
  \caption{\label{charfig} (Color online). Schematic illustration of
    shape of positive characterstics in the $x$-$y$ plane. In (a)
    $L_n$ is positive corresponding to diamagnetic drift opposing
    $E\wedge B$ drift, and the displacement $\delta_l$ being down the
    density gradient. In (b) $L_n$ is negative.}
  
\end{figure}

It is, in effect, the $\delta z$ variation that causes the positive
characteristics to have curvature. This curvature has magnitude
approximately $1/L_n$. When considering whether or not a probe surface
is convex, this curvature has to be accounted for. Schematic
representations of the positive characteristic shapes are illustrated
in Fig \ref{charfig}, for an object with two plane faces. When the curvature is
towards such a face, it is concave, and the plasma flows into the
probe with a negative parallel velocity greater than
$1-M_y\cot\theta$. This concavity is avoided in general if the probe
has a convex curvature that is greater than $1/L_n$. The
characteristics curve either away from or towards the line $x=const.$
depending on whether $L_n$ is positive or negative. ($M_y$ is taken
always positive.) 

To summarize, then, the ion flux per unit perpendicular area from the
plasma to a convex surface in the presence of combined (colinear)
$E\wedge B$ and density-gradient diamagnetic external drifts can be
written precisely as eq (\ref{flux}):
\begin{equation}\label{flux2}
  \Gamma_\parallel = n c_s = n_\infty(z_\infty) c_s \exp[-1 -(M_{\parallel\infty}-
    M_h\cot\theta)] ,
\end{equation}
 but with 
\begin{equation}
  M_h = M_E + M_{ni} 
\end{equation}
and 
\begin{equation}
  n_\infty(z_\infty) = n_\infty(z-\delta z) = n_\infty(z) \exp(-\delta z/L_n).
\end{equation}
Because of its near even parity in $M_\parallel+1$, the $\delta z$
term can generally be ignored for the purposes of $M_\perp$
determination. 

\section{Effects from presheath displacement}
\label{finite}
 
The expressions obtained so far are for the flux leaving the plasma
region where the drift expressions hold. Between that region and the
probe surface lie the magnetic presheath, of order a Larmor radius
thick, which we assume is small compared with the probe, and the Debye
sheath, of order 4 Debye lengths thick, which we shall ignore altogether.
The dynamics in the magnetic presheath are not ignorable. The flux to
the probe itself is different from the flux from the plasma into the
magnetic presheath when diamagnetic drifts are important.

\subsection{Magnetic presheath displacement calculation}

\begin{figure}[htp]
\includegraphics[width=.6\hsize]{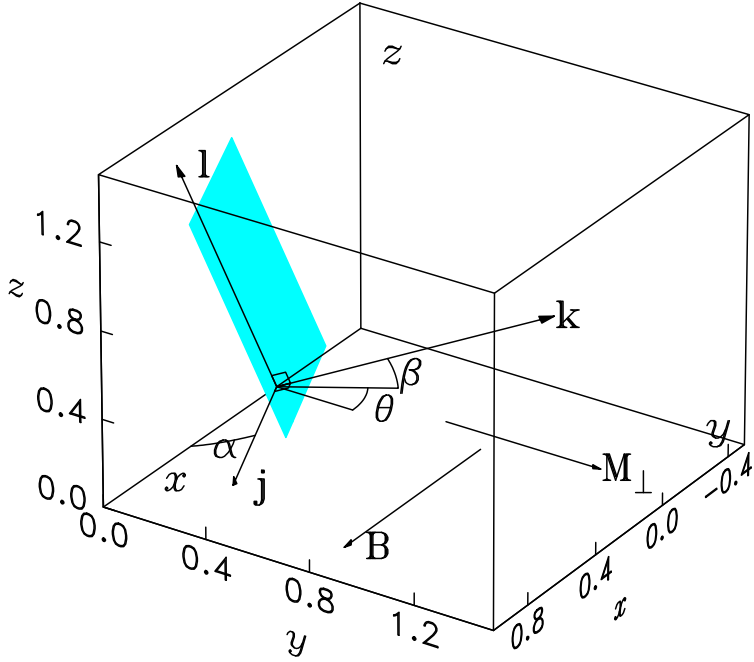}
\caption{\label{3dcoords}(Color online). Illustration of the coordinates
 referred to the field and drift directions ($x,y,z$) and the unit
  vectors referred to the
  probe surface directions ($\J,\K,\L$).
}
\end{figure}

In the magnetic presheath, the electric field normal to the surface is
strong enough that $E_\perp/B$ is of order the sound speed. The normal
gradients of the resulting drift give rise to non-negligible
convective derivative (i.e. polarization drift) terms. Indeed, it is
those terms that permit the ion fluid trajectory to acquire sufficient
cross-field velocity to satisfy the Bohm condition at the Debye-sheath
edge.  We assume on the magnetic presheath scale the probe surface can
be approximated as planar and the gradients can be considered all to
be normal to it.  Let the direction of the normal (outward from the
plasma) to the probe surface be $\K$. Define $\sin\alpha = -\K.\B/B$
and the unit vector $\L= \B\wedge\K/B\cos\alpha$ in the direction
perpendicular to $\B$ and $\K$. Define the third unit vector by
$\J=\K\wedge\L$. See figure \ref{3dcoords}.
Denote vector components in the
respective directions by subscripts. Then the (normal) $\K$-component
of the momentum equation (\ref{momneq}) is
\begin{equation}
  m \V.\nabla v_k = - Ze \nabla_k \phi - (1/n) \nabla_k p + Ze\V.(\B\wedge\K)
\end{equation}
we eliminate the potential
gradient using eq (\ref{phiint}), and density gradient using the
continuity equation (\ref{continuityeq}) in the form $nv_k=const.$, so that
$\nabla_k\ln n=-\nabla_k \ln v_k$, to obtain
\begin{equation}\label{mpsdrift}
  v_l \Omega_i\cos\alpha = (-c_s^2/v_k + v_k) \nabla_k v_k\ .
\end{equation}
We wish to calculate the total displacement, $\delta_l$ in the $\L$-direction,
experienced by the ion fluid traversing the magnetic presheath. This
is simply the time integral of eq (\ref{mpsdrift}):
\begin{equation}
  \delta_l \Omega_i\cos\alpha = \int (-c_s^2/v_k + v_k) \nabla_k v_k
  dt = \int (1 - c_s^2/v_k^2) dv_k= [v_k+c_s^2/v_k]\ ,
\end{equation}
(recognizing that $v_kdv_k/dx_k=dv_k/dt$). The limits of the integral
are $v_k/c_s=\Gamma_k/n c_s \equiv S$ at the magnetic presheath outer
edge (where $\Gamma_k$ is the normal flux density) and $v_k/c_s=1$ at
its inner edge, entering the Debye sheath (whose thickness we
ignore). Therefore
\begin{equation}
  \delta_l = \rho_s [2 -S - 1/S]/\cos\alpha
  = -\rho_s {[1-S]^2 \over S\cos\alpha}\ .
\end{equation}
Thus, for small incidence angle of the magnetic field ($S$ small) a
rather large displacement ($\delta_l\sim \rho_s/S$) along the
magnetic presheath occurs.
This displacement has been noted and roughly estimated
\cite{chankin94} in previous discussions of magnetic presheath
structure. Its significance is that ions exit the magnetic presheath
(and are collected by the probe) a tangential distance $\delta_l$ from
where they entered it. If there is a tangential gradient
$\nabla_l\Gamma_k$ of the normal flux-density $\Gamma_k$ entering the
magnetic presheath at position $x_l$, then the flux-density to the
probe will be not $\Gamma_k(x_l)$ but
$\Gamma_k(x_l-\delta_l)=\Gamma_k(x_l)\exp[-
  \delta_l\nabla_l\ln(\Gamma_k)]$ (choosing consistently with our $\ln
n$ assumption, uniformity of
$\nabla\ln\Gamma_k$ when surface curvature is ignored). This
alteration of the flux-density is precisely the phenomenon that Cohen
and Ryutov \cite{cohen95} calculated in the small-$\alpha$ limit. In
their Eulerian viewpoint, the alteration is attributable to divergence
of the tangential flux in the magnetic presheath. From the present Lagrangean
viewpoint, it arises from the integrated convective derivative. We can
demonstrate this by evaluation.
Identifying $\nabla\ln\Gamma_k = \nabla \ln n_\infty$ and using the
definitions of $S$ and $\L$ we find
\begin{eqnarray}\label{mpsdelta}
  \delta_l\nabla_l\ln\Gamma_k&=& -{[1-S]^2\over
    S\cos\alpha}\rho_s  {\L.\nabla \ln\Gamma_k }
= -{[1-S]^2\over S\cos^2\alpha\ c_s}   {T_i+ZT_e \over ZeB^2} (\nabla\ln
n_\infty\wedge \B).\K\nonumber\\
&=&  {[1-S]^2\over S\cos^2\alpha\ c_s}  (\V_{ni}-\V_{ne}).\K
=  {[1-S]^2 \hat{k}_y\over S\cos^2\alpha} (M_{ni}-M_{ne})
\end{eqnarray}
Cohen and Ryutov's calculation assuming small $S$ and $\alpha$ gave
this expression with the geometric term $[1-S]^2 \hat{k}_y /
\cos^2\alpha$ equal to 1, yielding a $\delta_l\nabla_l\Gamma_k$ equal
to the difference between the ion and electron diamagnetic normal flux
densities.  Now we evaluate the total flux density to the probe using
eq (\ref{flux2}), which yields $S = \sin \alpha$, and recognizing that
when the orientation of $\K$ perpendicular to $B$ is defined by
setting $\hat{k}_z=\sin\beta$, it immediately follows that $\hat{k}_x=
-\cos\beta\sin\theta$, $\hat{k}_y=\cos\beta\cos\theta$ and $\sin\alpha
= \cos\beta\sin\theta$. We obtain
\begin{equation}\label{generalflux}
  \Gamma_k(x_l-\delta_l) =
n c_s \sin\alpha \exp\left\{-\left[{1-\sin\alpha\over
    1+\sin\alpha}\right] (M_{ni}-M_{ne})\cot\theta\right\}\ .
\end{equation}
The ion flux  to the probe surface per unit perpendicular area is then 
\begin{equation}\label{flux3}
  \Gamma_{\parallel p} = n_\infty c_s \exp\left\{-1
  -M_{\parallel\infty} + M_h\cot\theta -\left[{1-\sin\alpha\over
      1+\sin\alpha}\right] (M_{ni}-M_{ne})\cot\theta\right\}\ .
\end{equation}
The logarithm of ratio of the flux for positive and negative
$\cos\theta$ is 
\begin{equation}
  \ln(\Gamma_+/\Gamma_-) = -{\delta z_+-\delta z_-\over L_n} + 2 M_h \cot\theta
  - 2{1-\sin\alpha\over 1+\sin\alpha} (M_{ni}-M_{ne})\cot\theta\ .
\end{equation}
Ignoring the $\delta z$ term, this becomes
\begin{equation}
   {\ln(\Gamma_+/\Gamma_-)\over 2\cot\theta} = M_E + { M_{ni}2\sin\alpha +
   M_{ne}(1-\sin\alpha)\over 1+ \sin\alpha} .
\end{equation}
Thus, in addition to $M_E$, the velocity combination measured by the
Mach probe ratio is an interpolation between the ion and electron
diamagnetic velocities (which of course have opposite sign), dependent
upon the angle $\alpha$ between the probe face and the field. Normally
in practice an intermediate value of $\theta$ (not too small but not too
close to $\pi/2$) must be used, which generally means an intermediate
value of $\alpha$.

\subsection{Probe curvature and finite facet size}

The displacement in the magnetic presheath also gives rise to a flux
correction when the probe surface has curvature. The flux at any point
on an electrode is characteristic of the flux into the magnetic
presheath at a position $-\delta_l$ away. If the surface is curved,
then the \emph{angles} at that position will be different, giving rise
to a different flux. But also, the displacement may have a divergence,
which gives rise to flux enhancement even if the flux into the
magnetic presheath were uniform. 

We note that the displacement strictly contains a component
$\delta_j$, along the direction $\J\equiv\K\wedge\L$, which for small
$\alpha$ is mostly along $\B$, and is of approximate magnitude
$\rho_s/\sin\alpha$, similar to $\delta_l$. This displacement
$\delta_j$ can be calculated in the form of a closed integral
expression by solving the magnetic presheath equations using the
techniques of references \cite{Chodura1982,riemann94}. We ignore it,
here and in the previous section, because it is mostly motion along
the magnetic field, which can be considered to be accounted for by
surface projection along the field, and because both the projection
and any additional cross-field motion (in the $\K$-direction) that
makes $\delta_j$ different from the pure field-line projection, make
flux contributions that are symmetric: of even parity under reversals
of $k_y$ and hence of
$\cos\theta$. In other words, there is a small correction to the total
flux density from $\delta_j$, which can be pictured as arising from
the fact that the probe collects ions from a cross-field area that is
larger than its solid cross-section by a margin of width approximately
$\rho_s$. But that correction does not affect the deduced
Mach-numbers, because they are based on ratios of collection fluxes
from surfaces with opposite $k_y$, which are equally perturbed by
even-parity terms.

 The
perpendicular flux-density perturbation arising from $\delta_l$ along
$\L$ has contributions $\delta_l\partial n/\partial x_l= \delta_l n M_h
\partial\cot\theta/\partial x_l$ and from the convective derivative of
the perpendicular area element, which can be written (differentially)
 $\Delta A / A = \nabla^s(\delta_l \L)= \delta_l \nabla^s.\L + \L.\nabla^s \delta_l$,
where $\nabla^s$ denotes the two-dimensional gradient
$(\partial/\partial y, \partial/\partial z)$ in the perpendicular
coordinates, but evaluated along the probe surface (not at constant $x$).
Thus the total correction arising from surface curvature is 
\begin{equation}
 - \Delta\Gamma_\parallel/\Gamma_\parallel = M_h \delta_l \L.\nabla^s
  \cot\theta + \delta_l \nabla^s.\L + \L.\nabla^s \delta_l \ .
\end{equation}
In this equation the first term has even parity with respect to
$k_y$-reversal and therefore does not contribute to flow-measurement
bias. The last two have odd parity and do contribute. Their order of
magnitude is $\delta_l/R_c$ where $R_c$ is the typical radius of
curvature of the surface. (Note that this is a real physical effect,
not the elementary mathematical integration discussed by Pelemann et
al\cite{peleman06b}.)

\begin{figure}[ht]
  \includegraphics[width=0.5\hsize]{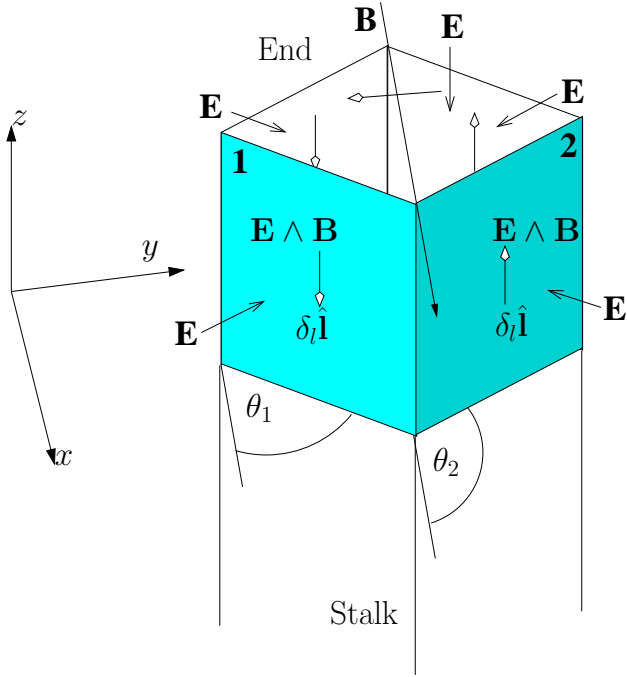}
  \caption{\label{endeffect}(Color online). Illustration of the
    opposite directions of magnetic presheath displacement ($\delta_l
    \L$) for opposite values of $\cos\theta$. The electric field is
    towards the probe and $\delta_l \L$ is in the $\E\wedge\B$
    direction. When the displacement is towards the stalk, as it is
    for facet 1 but not facet 2, a depleted region exists at the probe
    end. (In this figure ``forward'' is up and ``backward'' is down.)}
\end{figure}

A related, but more intuitive and probably more relevant correction
comes from edges of plane facets. Most Mach probes have electrodes at
or near their ends \cite{MacLatchy1992,gunn01,gangadhara04,shikama05}
The direction $\delta_l\L$ associated with the surfaces points
forward, beyond the probe end, on one side and backward towards the
probe stalk on the other, as illustrated in Fig \ref{endeffect}. The
surface for which it points backward collects essentially zero flux
for a distance $\approx\delta_l$ from the end, because of the magnetic
presheath displacement; then it collects the full flux from then
on. The surface for which $\delta_l\L$ points forward collects full
flux at its end. The electrode will collect full flux throughout its
area, provided that it is embedded in a surface of constant angle
which extends a distance $\approx\delta_l$ past the (backward) edge of
the electrode.  Therefore the likely effect (dependent on the detailed
electrode placement) of $\delta_l$ is to induce a depleted flux region
on \textit{just the forward edge} of \textit{just the electrode for
  which $\delta_l\L$ is backward}. An appropriate estimate of the
odd-parity part of the resulting flux change is
\begin{equation}
  \Delta\Gamma_\parallel/\Gamma_\parallel \approx  \delta_l
  \hat{l}_z/2\bar{h} = - {\rho_s(1-S)^2\cos\beta\over2\bar{h} S\cos\alpha  }
\approx -\rho_s{(1-\sin\alpha)^2 \over  2\bar{h}\cos\alpha
  \sin\theta} \ ,
\end{equation}
approximating $S$ as $\sin\alpha$ to lowest order in $M_\perp$, and
denoting the average $z$-extent of the electrode as $\bar{h}$. 

Since in a perpendicular Mach-probe measurement $M_h$ is estimated
from the expression 
\begin{equation}
  M_h |\cot\theta| = (1/2)\ln[
    \Gamma_{\parallel+}/\Gamma_{\parallel-}]\ ,
\end{equation}
(where subscript $\pm$ refers to the sign of $\cos\theta$)
the perturbation $\Delta M_h$ to the deduced $M_h$ is related to the odd-parity
perturbation to flux, $\Delta\Gamma_\parallel$, by 
\begin{equation}
  \Delta M_h = |\tan\theta| \Delta \Gamma_\parallel /\Gamma_\parallel
  \approx  -{\rho_s\over 2\bar{h}} {(1-\sin\alpha)^2 \over \cos\alpha
  \cos\theta}\ .
\end{equation}
There is therefore an intrinsic bias of order $\rho_s/2\bar{h}$ in
such a Mach-probe measurement. Its direction is such that if the
gradient of plasma pressure is in the forward direction of the probe,
in other words the probe is introduced from the ``outside'' of the
plasma where pressure is low (as is generally the case), then the
spurious apparent drift is in the same direction as the
ion-diamagnetic drift.  This end-effect can in principle be avoided if
the electrode does not sample the plasma at the end of the facets, but
instead approaches the end no closer than a $z$-distance of
approximately $\rho_s(1-S)^2\cos\beta/(S\cos\alpha)$.

\section{Temperature gradient drifts}

Diamagnetic drifts might also arise from temperature gradients, which
have so far been excluded. A physically justifiable approach for
electron temperature gradients simply regards $T_e$ as constant along
the field, but having an externally imposed gradient in the
$z$-direction. A similar mathematical ansatz will be applied to $T_i$
but with less clear physical justification. The equations of
continuity and parallel momentum (\ref{parmomen}) may once more be
rearranged into characteristic form to obtain
\begin{equation}\label{gradtcharact}
  (\V.\nabla\pm c_s\nabla)[\ln n/n_\infty \pm v_\parallel/c_s] =
\pm (v_\parallel/c_s)(-v_z/L_c)- v_z/L_p\ .
\end{equation}
Here there are two inhomogeneous terms on the RHS. The first arises
from the variation of $c_s$ with $z$ because of temperature
gradients. Its gradient has been written as $dc_s/dz=c_s/L_c$. The
second arises because of (a) the gradient of $n_\infty$ (compare with
eq (\ref{ncharacteristics})) and (b) the divergence $\V_\perp$
acquires when there are temperature gradients, which can be written
$\nabla.\V_\perp= v_z/L_T$, where $L_T\equiv T_i/(dT_i/dz)$ is the
scale-length of ion-temperature gradient. Because (a) and (b) are of
identical form, we combine them as $-v_z/L_p$ with $1/L_p \equiv 1/L_T
+ 1/L_n$.

The inhomogeneous terms on the RHS are the same order as the LHS terms
near the boundary of the unperturbed region. They cannot therefore
simply be ignored because of ordering. Their effects are to change the
value of $\ln n/n_\infty - {v_\parallel/c_s}$ along the negative
characteristic, which is what determines the flux at the
boundary. This change can be calculated, and proves to be usually
ignorable, as we now show. We assume that the value of this negative
characteristic quantity can be expressed as a function only of
$M_\parallel$:
\begin{equation}\label{gdef}
  \ln n/n_\infty - M_\parallel = g(M_\parallel)
\end{equation}
and solve for $g$ using the same approach as we used for density
gradients. In outline,
the calculation is as follows.

The positive characteristic equation gives
\begin{equation}
-v_z \left({M_\parallel \over L_c}+{1\over L_p}\right) = \ddtpn (2M_\parallel +g) = \ddtpn (2\ln
n/n_\infty -g)  
\end{equation}
We define $dg/dM_\parallel=2q$ so that ${\left.d\over dt\right|_\pm} g=2q{\left.d\over dt\right|_\pm} M_\parallel$
and eliminate $M_\parallel$ derivatives from the characteristics to
obtain
\begin{equation}\label{derivs}
  \ddtmn g = -v_z \left({-M_\parallel \over L_c}+{1\over L_p}\right)\quad,
\quad \ddtpn g= -v_z \left({M_\parallel \over L_c}+{1\over L_p}\right){q\over 1+q} . 
\end{equation}
We eliminate partial $x$-derivatives between the
positive and negative characteristics:
\begin{equation}\label{combchar}
  -(M_\parallel - 1)\ddtpn g + (M_\parallel + 1) \ddtmn g= 2 c_s M_h
  {\partial\over \partial y}g \ .
\end{equation}
We re-express $y$-derivatives
using the $z$-velocity expression (\ref{grady}), which yields
\begin{equation}\label{dgdy}
  M_z = \rho_s(1+1/2q)\partial g/\partial y .
\end{equation}
Substituting (\ref{derivs}) and (\ref{dgdy}) into (\ref{combchar}) and dividing the resultant through
by $v_z$ we arrive at
\begin{equation}
-(M_\parallel -1)\left({M_\parallel \over L_c}+{1\over L_p}\right) {q\over 1+q}+
(M_\parallel +1)\left({-M_\parallel \over L_c}+{1\over L_p}\right) =
{-2 M_h \over \rho_s(1+1/2q)} \ .
\end{equation}
This quadratic equation for q can be solved to obtain
\begin{equation}\label{quadcs}
  {1\over 2}{dg\over dM_\parallel} = q = 
{2(M_\parallel^2-s-w)+(1-s)M_\parallel\pm\sqrt{[(1-s)^2-4]M_\parallel^2+2(s+2)w}
\over
-4(M_\parallel^2 -s-w)
}
\end{equation}
where $w\equiv M_hL_c/\rho_s$ and $s\equiv L_c/L_p$. 

Provided $M_{\parallel\infty}$, $w$, and $s$ are indeed constant, this is
a consistent solution. In view of the variation of $c_s$ and hence
$\rho_s$ with $z$, it is clear that the constancy of $u$ requires
$c_s(z)$ to satisfy a simple differential equation, whose solution
gives the required $c_s(z)$. This is not as simple a profile as the
linear $\ln n_\infty$ profile for the density-gradient. But we are
interested in a local region on the gradient, where it is acceptable
to choose the precise form of $c_s(z)$ to satisfy the consistency
condition without substantially changing the problem, in-so-far as a
local expansion of $c_s$ is allowable.

\begin{figure}[htp]
  \includegraphics[width=0.6\hsize]{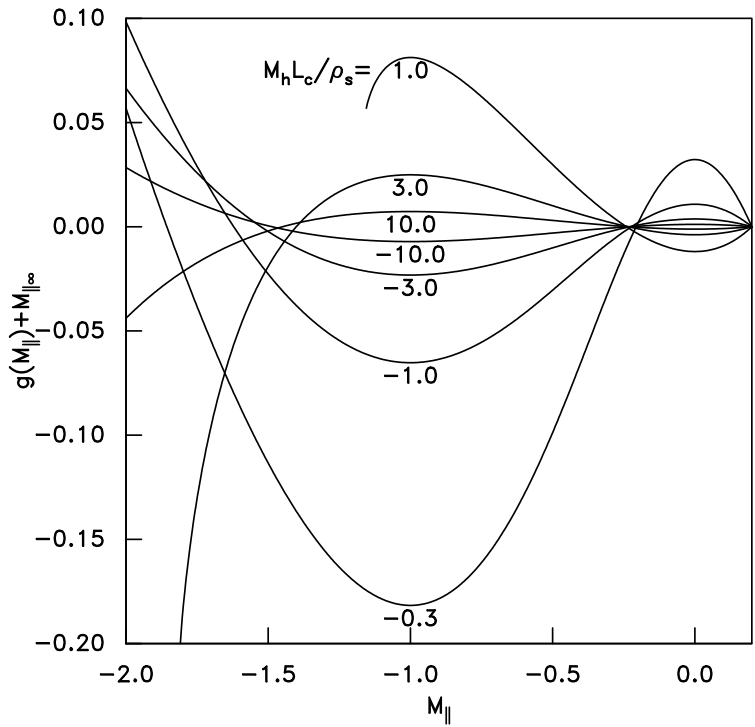}
  \caption{\label{gofm}Solutions for the quantity $g+M_{\parallel\infty}=\ln n/n_\infty
    -M_\parallel+M_{\parallel\infty}$ accounting solely for perpendicular gradients of
    sound-speed ($s=0$). }
\end{figure}

The limit of eq (\ref{quadcs}), when $s$ and $w \to \infty$ together,
is minus eq (\ref{requ}) although with different notation.  Equation
(\ref{quadcs}), can also be integrated analytically, but the resulting
expression is long and cumbersome, so it is not given here. Instead,
numerical integrations are illustrated in Fig \ref{gofm}. We choose to
illustrate $s=0$, which corresponds to a case where only the electron
temperature has an external gradient.
The case $M_{\parallel\infty}=0.2$ is shown, to emphasize the
applicability of the solution shapes to all $M_{\parallel\infty}$. The
solution becomes imaginary for some relevant negative values of
$M_\parallel$ when $0<w\lesim 7$, illustrated by the $w=1,3$
curves. Provided such parameters are avoided, it can be seen that the
values and especially the odd parity (in $M_\parallel+1$) part of
$g+M_{\parallel\infty}$ are small.

A second modification of the governing equations, arising from
temperature gradients, is that additional terms are present in the
drift velocity eq (\ref{inclpol}). Ion temperature gradients are easy
to treat because they can be expressed as a uniform term like $\nabla
\phi_\infty$, but $T_e$ gradients introduce a new type of term
(c.f. \ref{gradndrift}). The total drift velocity is:
\begin{equation}
     \V_\perp= -\left[\ln(n/n_\infty)\nabla \left(T_e\over e\right)
       +\nabla \left(\phi_\infty+ {T_i\over Ze}\right) + {T_i\over Z
         e}\nabla\ln n_\infty + {m c_s^2\over Ze} \nabla \ln
       n/n_\infty\right] \wedge{\B\over B^2}\ .
\end{equation}
We can still discard the final ($\nabla\ln n/n_\infty$) term as an
$\M_{\perp1}$ giving no advection because $\M_{\perp1}.\nabla \ln
n/n_\infty=0$. The rest should be regarded as $v_h$, having no
$z$-component. The first term is the new one, so far not treated. It
is non-uniform; but a function of $\ln n/n_\infty$ and hence, within
the solution schemes we have developed, a function of
$M_\parallel$. The $v_h$ variation can be incorporated into the scheme
above, except that the consistency condition upon the $c_s$ profile
becomes that $w\ (=M_hL_c/\rho_s)$, which varies with $v_h$, must
be a function only of $M_\parallel$. Assuming an appropriate $c_s$
shape is chosen, eq (\ref{quadcs}) still applies, and could still be
integrated provided that $w(M_\parallel)$ were known. Plainly,
provided that $w$ stays within acceptable ranges during the
integration, curves that bear resemblance to those of Fig (\ref{gofm})
will be obtained. Generally then, the arguments that
$g+M_{\parallel\infty}$ can usually be ignored will remain as valid
as before, provided $w$ stays within the range of their validity.

The
boundary condition at the plasma edge is still that the positive
characteristic be tangent to the probe surface, that is 
(eq \ref{curved}) $M_\parallel = M_h \cot\theta -1$, but with the
following definition:
\begin{eqnarray}
  \M_h &=& -\left[\ln(n/n_\infty)\nabla \left(T_e\over e\right) +\nabla
    \left({T_i\over Ze}+\phi_\infty\right) + {T_i\over Z e}\nabla\ln
    n_\infty \right] \wedge{\B\over B^2c_s}\\
\label{tdrift}
   &=& -\ln(n/n_\infty)M_{Te} + M_{Ti}+ M_E + M_{ni}\ ,
\end{eqnarray}
where $M_{Te}$ and $M_{Ti}$ are the (external) diamagnetic drift
due to temperature gradient of the electrons and ions.
We can then substitute into eq 
(\ref{gdef}) to eliminate $M_\parallel$ and obtain after rearrangement:
\begin{equation}
  \ln n/n_\infty = { -1 + g + (M_{Ti}+M_E+M_{ni})\cot\theta \over
    1 + M_{Te}\cot\theta}\ ,
\end{equation}
in which, for the purposes of perpendicular velocity measurement, we
can take $g\approx-M_{\parallel\infty}$. To first order in the
perpendicular velocities this can be written
\begin{equation}
  \ln n/n_\infty \approx -1 -M_{\parallel\infty} +
      [(1+M_{\parallel\infty})M_{Te}+M_{Ti}+M_E+M_{ni}]\cot\theta \ ,
\end{equation}
showing that the effect of the electron temperature-gradient drift can
be amplified or attenuated, compared with the other drifts, depending
upon the parallel external velocity $M_{\parallel\infty}$.

Finally we must account for the temperature gradients in the
correction arising from magnetic presheath displacement. We can ignore
the next order corrections to $\delta_l$ arising from approximations
in its derivation. But we must account for the fact that since
$\delta_l \propto c_s$, a transverse derivative of $c_s$ gives rise to
transverse divergence which alters the flux density as it traverses
the magnetic presheath. The change in area $A$ arising from this
effect is $\Delta A/A = \delta_l \L.\nabla \ln c_s$. This gives rise
to a change in $\ln \Gamma$ of $-\delta_l \L.\nabla \ln c_s$. In
addition, the displacement gives rise to a convective difference
between the flux to the probe and that entering the presheath that,
since $\Gamma \propto n c_s$ can be expressed as
\begin{equation}
  \Delta \ln(nc_s) = - \delta_l \L.(\nabla \ln n +
  \nabla \ln c_s) 
\end{equation}
adding these two effects gives the total magnetic presheath difference:
\begin{equation}
  \Delta \ln \Gamma_k = - \delta_l \L.(\nabla \ln n +
  2 \nabla \ln c_s) = - \delta_l \L.(\nabla \ln n +
  \nabla \ln(ZT_e+T_i)) 
\end{equation}
Applying the transformations of eq (\ref{mpsdelta}), we see that our
result is the same as before except that the total diamagnetic
difference velocity $M_D\equiv M_{Di}-M_{De}=(M_{ni}+M_{Ti})-(M_{ne}+M_{Te})$ is
involved rather than just the density-gradient part calculated
before. 

The final expression for flux per unit perpendicular area to the probe
is thus
\begin{equation}\label{final}
  {\Gamma_{\parallel p} \over nc_s} = \exp\left\{
{ -1 + g + (M_{Ti}+M_E+M_{ni})\cot\theta \over
    1 + M_{Te}\cot\theta} 
-\left[{1-\sin\alpha\over
      1+\sin\alpha}\right] (M_{Di}-M_{De})\cot\theta
\right\},
\end{equation}
or, to first order in perpendicular velocities, and approximating $g$,
\begin{equation}\label{finalap}
 \ln \left\{\Gamma_{\parallel p} \over nc_s\right\} = 
 -1 -M_{\parallel\infty} +
      \left[(1+M_{\parallel\infty})M_{Te}+M_{Di}+M_E
-\left({1-\sin\alpha\over
      1+\sin\alpha}\right)M_D
\right]\cot\theta ,
\end{equation}
where $\alpha$ is the angle between the probe surface and the magnetic
field (in 3-dimensions), and $\theta$ is the angle within the plane containing
field and external drift. All drift velocities here refer to the
external, unperturbed plasma.

\section{Discussion}

The complete solution of the drift equations that has been obtained
here is highly appropriate when the $E\wedge B$ drift dominates. Then
the positive characteristics are straight and the full solution in the
plasma region can readily be constructed. Notice that the parallel
length of the presheath (i.e. the perturbed plasma region) is
approximately the transverse size of the probe $r_p$ (say) times
$(M_\parallel+1)/M_\perp$. If the perpendicular Mach number is not
very small, this length $\sim r_p/M_\perp$ is likely substantially
shorter than that from the standard diffusive estimate: $r_p^2
c_s/D$. Certainly it can easily
be shorter than the mean-free-path for electron-ion Coulomb
collisions, which is required for the ignoring of friction inherent in
our treatment. In such a situation taking the electron temperature to
be invariant along the field is completely natural. It is not so
natural to make that approximation for the ions. However, it is known
from other calculations \cite{Chung1988,gunn01} that the isothermal
approximation gives results quite close to those that arise from more
physically plausible approximations. In any case, the standard
widely-accepted formulas are based upon isothermal-ion calculations.
In addition to providing rigorous analytical justification, regardless
of probe geometry, for formulas that are practically the same as those
arising from a diffusive treatment, the present treatment helps to
resolve another conceptual problem of long standing. It is that probes
are often smaller than the typical transverse size of the turbulence
that is responsible for transport in the regions of plasma in which
they are used. In other words, transport, for example in tokamak edges
and scrape-off-layers, is actually known to be dominated by
fluctuating cross-field flows that in many situations have eddies
larger than the probes. In such situations, using a heuristic
approximation that cross-field flux is expressible through a diffusion
coefficient, as prior treatments have done, is questionable. The
present treatment, regarded as a short-time snap-shot of a situation
that is fluctuating, is more appropriate. Fortunately the result is
the same, although with much sharper physical justification.

The drift-approximation is unproblematic for $E\wedge B$ drifts,
because there is nothing to prevent the probe being bigger than the
Larmor radius while the drift Mach number is of order unity. This is
not so for diamagnetic drifts. The diamagnetic Mach number is of order
$M_D\sim\rho_s/L$ (where $L$ is the pressure
scale-length). Consequently if $M_D\sim 1$, there is no separation of
scales between the Larmor radius and the gradient scale-length. It is
then impossible to choose a probe size that is both bigger than the
Larmor radius (in order to justify the drift approximation) and
smaller than the gradient scale-length (to justify a local
approximation of the drifts). This is an inherent difficulty that
underlies the need in the calculations to choose a specific shape of
the plasma profiles. Only if $M_D$ is small does a local approximation
to the flow make sense. Therefore, the usefulness of the present
calculation is increasingly compromised as $M_D$ approaches unity.
A calculation that avoids the drift-approximation is then really
needed. It seems unlikely that an analytic solution like the present
one will be forthcoming.

In summary, complete solutions of the problem of ion collection by an
arbitrary-shaped object have been obtained in the drift approximation
(ignoring the polarization drift).  The normalized flux density
(\ref{final}), (\ref{finalap}) is a function only of the orientation
of the surface, provided the object is convex. The precise meaning of
``convex'' here is that no positive characteristic that originates
elsewhere on the object should pass through
the point of interest. A transverse Mach probe using a
variety of electrode orientations measures the $E\wedge B$ drift plus
this specified combination of both ion and electron diamagnetic drifts.
Its ideal calibration has been derived, and possible problems arising
from finite size identified and quantified.

\bibliography{hutch}

\end{document}